\documentclass[twoside,twocolumn,english,superscriptaddress,nofootinbib,preprintnumbers, aps]{revtex4}
\usepackage{color}
\usepackage{amsmath}
\usepackage{graphicx}
\usepackage{amssymb}
\usepackage{dsfont}
\usepackage{soul}
\usepackage{esint}
\usepackage[hyperfootnotes=true,unicode=true, 
 bookmarks=true,bookmarksnumbered=false,bookmarksopen=false,
 breaklinks=false,pdfborder={0 0 1},backref=false,colorlinks=true]
 {hyperref}
 \usepackage{float}
 \usepackage[markup=nocolor]{changes}%resoinsible for sout

\makeatletter
\@ifundefined{textcolor}{}
{%
 \definecolor{BLACK}{gray}{0}
 \definecolor{WHITE}{gray}{1}
 \definecolor{RED}{rgb}{1,0,0}
 \definecolor{GREEN}{rgb}{0,1,0}
 \definecolor{BLUE}{rgb}{0,0,1}
 \definecolor{CYAN}{cmyk}{1,0,0,0}
 \definecolor{MAGENTA}{cmyk}{0,1,0,0}
 \definecolor{YELLOW}{cmyk}{0,0,1,0}
 }

\usepackage{hyperref}
\hypersetup{
    colorlinks,%
    citecolor=blue,%
    filecolor=blue,%
    linkcolor=blue,%
    urlcolor=blue
}

\makeatother

\begin{document}

\title{Cosmological Effective Hamiltonian from full Loop Quantum Gravity Dynamics}

\author{Andrea Dapor}
\email{andrea.dapor@gravity.fau.de}
\affiliation{Institute for Quantum Gravity, Friedrich-Alexander University Erlangen-N\"urnberg, Staudstra\ss e 7, 91058 Erlangen, Germany}

\author{Klaus Liegener}
\email{klaus.liegener@gravity.fau.de}
\affiliation{Institute for Quantum Gravity, Friedrich-Alexander University Erlangen-N\"urnberg, Staudstra\ss e 7, 91058 Erlangen, Germany}

\date{\today{}}

\begin{abstract}
The concept of effective dynamics has proven successful in LQC, a loop-inspired quantization of cosmological spacetimes. We apply the same idea of its derivation in LQC to the full theory, by computing the expectation value of the scalar constraint with respect to some coherent states peaked on the phase-space variables of flat Robertson-Walker spacetime. We comment on the relation with effective LQC and find a deviation stemming from the Lorentzian part of the Hamiltonian.
\end{abstract}

\maketitle
\section{Introduction}
With new data being collected each year, the era of observational cosmology is at its peak. Oberservations of the CMB make it possible for the first time to study the very early Universe and Planck-scale physics. By this we can test certain models of quantum cosmology \cite{BCT11}, however there remain yet unanswered questions at the deep core of quantum gravity. Thus one must finally find ways to connect cosmology with these models, which make predictions about the fine structure of quantum space-time.

A promising candidate, Loop Quantum Gravity (LQG), has over the last decades evolved into a rich and mathematically well-defined theory of quantum gravity. While its backbone has been set up through a lot of work in the framework of the full theory \cite{Rov04,Thi07,GS13}, the same procedure of loop-quantisation applied to the symmetry-reduced sector has pushed the field into the direction of observable predictions and managed to resolve some of the open questions from the classical theory. For example, in Loop Quantum Cosmology (LQC) the Big Bang singularity is resolved by the ``big bounce scenario'' \cite{Boj08,AS11,AS16}.

In this approach one deals with states representing 3-dimensional geometries: the data on a spatial Cauchy slice of the spacetime. While different choices of such slices do not affect physical quantities per se, depending on the system there exist convenient choices. This amounts to performing a gauge-fixing of the constraints via some degrees of freedom (usually provided by matter), which therefore play the role of physical clocks and rods (however depending on the matter choice the observables may differ) \cite{GT15,GO17}.

At this point, the theory is reduced to a (infinite-dimensional) version of quantum mechanics: a quantum 3-geometry is a physical state, $|\psi\rangle$, and its evolution (wrt the physical time $\phi$) is given by a physical Hamiltonian $\hat H$ (derived from the scalar constraint) by
\begin{align} \label{time-evol-psi}
|\psi(\phi)\rangle = e^{-i\phi\hat H} |\psi(0)\rangle
\end{align}
One can then ask how quantities of interest (e.g., geometrical operators such as volume $\hat V$ of the universe) change in time:
\begin{align} \label{time-evol-V}
V_\psi(\phi) := \langle \psi(\phi) | \hat V | \psi(\phi) \rangle
\end{align}

Computations such as (\ref{time-evol-V}) are numerically possible in the context of LQC, and in the seminal papers \cite{APS-PRL,APS} it was shown that the quantum evolution of the expectation value of observables (in particular, the volume of the universe) on certain coherent states labelled by loop variables $(p,c)$ follows closely some ``quantum-corrected'' trajectories in phase space.\footnote
{
The relation between $(p,c)$ and the Ashtekar-Barbero variables is obtained by fixing a fiducial Minkowski metric. One then finds that the connection and the densitized triad are given by
\begin{align}
A_a^I = c \delta^I_a, \ \ \ \ \ \ \ E_I^a = p \delta^a_I
\end{align}
where we set the coordinate volume of the universe to $1$ for simiplicity. The variables $c$ and $p$, which will depend on the (cosmological) time, thus encode all the dynamics of the model.
}
These trajectories correspond to the integral curves of an effective Hamiltonian. Such Hamiltonian is not the classical cosmological one -- which in terms of $(p,c)$ reads
\begin{align} \label{classical-rw-ham}
H_{\text{cl}}(p,c) = -\dfrac{6}{\kappa \beta^2} \sqrt p c^2
\end{align}
(with $\kappa = 16\pi G/c^3$ the gravitational coupling constant and $\beta$ the Immirzi parameter) -- but rather the phase-space function obtained by taking the expectation value of the quantum Hamiltonian $\hat H_{LQC}$ on a semiclassical coherent state, $\psi_{(p,c)}$:
\begin{align} \label{effective-rw-ham}
H_{\text{eff}}(p,c) := \langle \psi_{(p,c)} | \hat H_{LQC} | \psi_{(p,c)} \rangle = -\dfrac{6}{\kappa \beta^2} \sqrt p \dfrac{\sin^2(c\mu)}{\mu^2}
\end{align}
where $\mu$ is a parameter of the quantum theory. The authors of \cite{APS} refer to this as the ``effective Hamiltonian'', and to the evolution it produces (which is equivalent to a corrected version of Friedmann equations) as ``effective dynamics''. In particular, it is possible to show that $H_{\text{eff}}$ bridges a contracting universe in the far past (which would evolve towards a Big Crunch according to $H_{\text{cl}}$) with an expanding branch in the far future (which would come from a Big Bang according to $H_{\text{cl}}$) via an intermediate ``bouncing'' region during which the energy density reaches the Planckian regime \cite{APS}. The cosmological singularity is then averted by this so-called ``big bounce''.
\section{Hamiltonian Operator in LQG}
 Inspired by this success, one may ask whether the same qualitative behavior is also found in the full version of the theory. In LQC the effective Hamiltonian can be obtained by computing the expectation value of $\hat{H}_{LQC}$ with respect to some coherent states \cite{Tav08}. In order to see whether this could also work for full LQG, the first step of this programme is to compute an equivalent expectation value there. As we are considering the quantum version of a field theory (general relativity) one starts as in Lattice gauge theories by introducing a UV cutoff. This restricts the infinite dimensional degrees of freedom to a cubic graph $\gamma$. We embed $\gamma$ such that its 3 directions to define the 3 axes of coordinates of a compact manifold $\sigma$ with periodic boundary conditions, i.e., a $3$-Torus. Hence $\gamma$ has a finite number of vertices, $\mathcal{N}^3$, and setting its coordinate volume to $1$ we find that the coordinate distance between two nearby vertices is $\mu := 1/\mathcal N$. 
As we have rewritten general relativity using a $3+1$ split of our manifold, it turns out to be useful to pass form the spatial metric $q_{ab}$ to ($A^I_a, E^a_I$), the so called Ashtekar-Barbero variables \cite{Ash87,Ash88,Bar94,Bar95,Bar96}. An advantage of this canonical pair is that, upon quantization, the Hilbert space over every edge $e\in \gamma$ can be expressed as $\mathcal{H}_e = L_2(SU(2),d\mu_{H})$ with $d\mu_{H}$ being the Haar-measure. Like in gauge theories, we associate with the holonomy of the connection $A^I_a$ a multiplication operator and a derivative operator for its canonical counterpart $E^a_I$:
\begin{align}\label{hol-op}
\hat{h}_{mn}(e) f_{e}(g) = D^{(\frac{1}{2})}_{mn}(g) f_{e}(g)
\end{align}
and
\begin{align}\label{flux-op}
-\frac{4}{i\hbar\kappa\beta}\hat{E}^k(e) f_{e}(g) = R^k f_{e}(g):=\frac{d}{ds}\mid_{s=0} f_{e}(e^{-is\sigma_k}g)
\end{align}
where $D^{(\frac{1}{2})}_{mn}$ is the Wigner matrix of group element $g$ in spin-$1/2$ $SU(2)$-irrep, $R^k$ is the right-invariant vector field, and $\sigma_k$ are the Pauli matrices.

In the mentioned framework of the Ashtekar-Barbero variables the classical scalar constraint $H$ can be written as
\begin{align} \label{classic-fullham}
H =& H_E-(\beta^2+1)H_L
\end{align}
where
\begin{align}
& H_E=F_{ab}^I\epsilon_{IJK}\frac{E^a_JE^b_K}{\sqrt{|\text{det}(E)|}}
\\
& H_L=\epsilon_{IMN}K^M_aK^N_b\epsilon^{IJK}\frac{E^a_JE^b_K}{\sqrt{|\text{det}(E)|}}
\end{align}
are called the Euclidean and Lorentzian part respectively. Here, $F$ denotes the Lie algebra valued curvature of connection $A$. Now, Dirac quantization scheme is employed to promote the function $H$ to an operator $\hat H$ on $\mathcal{H} := \otimes_e \mathcal{H}_e$, whose details depend on the choice of regularization. Here, we shall focus on the proposal \cite{GT06-2}, i.e., a {\it non-graph-changing} operator which uses the regularization proposed by Thiemann in \cite{Thi96_1,Thi96_2}. There, the fundamental observation is the classical equality
\begin{align}
K^I_a = \frac{2}{\kappa\beta^3}\{A^I_a, \{H_E,V\}\}
\end{align}
This allows to quantize (\ref{classic-fullham}) in terms of (\ref{hol-op}), (\ref{flux-op}) and the Ashtekar-Lewandowski volume operator $\hat V$ \cite{AL97,GT_CC06}.
\section{Coherent States and Volume}
To follow the program of effective dynamics, we now must choose a set of coherent states $\Psi_ {(A,E)} \in \mathcal H$ and evaluate the expectation value of $\hat H$ on them. A choice which is peaked on {\it both} holonomies and fluxes is the \emph{complexifier coherent states}, developed by Thiemann and Winkler \cite{Thi00_I,TW01_II, TW01_III, Thi02} on the basis of Hall's work \cite{Hall94,Hall97}. These are labelled on every edge by $h_e\in\text{SL}(2,\mathbb{C})$, which can be written in the holomorphic decomposition as:
\begin{align}\label{holomorphic-dec}
h_e = n_e e^{-i\bar{z}_e\sigma_3/2} {n'_e}^{\dagger}
\end{align}
with $n_e,n'_e\in \text{SU}(2)$ and $z_e\in\mathbb{C}$. Then, we can write explicitly for the coherent state $\Psi_{(A,E)} := \bigotimes_{e\in\gamma} \psi_{e,h_e}$ where
\begin{align}\label{coherent-state}
\psi_{e,h_e}(g)&=\frac{1}{N}\sum_{j=0}^{\infty}d_je^{-j(j+1)t} \sum_{m=-j}^j e^{izm}D^{(j)}_{mm}({n_e}^{\dagger} g n'_e)
\end{align}
with $N^2=||\psi_{e,h_e}||^2$ the normalization of the state and $d_j = 2j +1$ the dimension of spin-$j$ $SU(2)$-irrep. The dimensionless quantity $t\in\mathbb{R}^+$ is the {\it semiclassicality parameter}.\footnote
{
For reasons to be discussed in an extended companion paper, $t$ should be identified with a ratio between areas:
\begin{align} \label{t-meaning}
t = \dfrac{\ell_p^2}{a^2}
\end{align}
where $\ell_p^2=\hbar\kappa$ is Planck length and $a$ is another scale of units of length that the theory should provide. For example, if $\Lambda$ is the cosmological constant, we can set $a = \Lambda^{-\frac{1}{2}}$, so we find $t = \ell_p^2 \Lambda \sim 10^{-120}$.
}

The choice of these coherent states is not only justified by the fact that they are peaked in the elementary operators, but also that it simplifies the analysis of the quantum Hamiltonian: as shown in \cite{GT06}, when considering expectation values on complexifier coherent states, the Ashtekar-Lewandowski volume operator can be arbitrarily approximated to order $\mathcal O(t^{k+1})$ by the Giesel-Thiemann volume operator
\begin{align} \label{volume-expansion}
& \hat{V}^{GT}_{k,v} := \langle  \hat Q_v  \rangle^{1/2} \sum_{n = 0}^{2k+1} c_{k,n} \frac{\hat{Q}_v^{2n}}{\langle\hat Q_v \rangle^{2n}}
\end{align}
where $c_{k,n}$ are known coefficients and $\langle \hat Q_v \rangle$ is a shorthand for the expectation value on coherent state $\Psi_{(A,E)}$ of the operator
\begin{align}\label{al-volume}
\hat{Q}_v=i \frac{(\beta\hbar\kappa)^{3}}{2^{10} 3} \sum_{e\cap e'\cap e''=v}\epsilon(e,e',e'')\epsilon_{ijk}\hat{R}^i(e)\hat{R}^j(e')\hat{R}^k(e'')
\end{align}
with $\epsilon(e,e',e'') := \text{sgn}(\det(\dot e,\dot e',\dot e''))$. Thanks to this result, $\hat H$ becomes a computable operator and we retain control on the error we make in terms of powers of the semiclassicality parameter $t$.
\section{The Hamiltonian}
We can now compute the expectation value of the Hamiltonian of the full theory (on a cubic graph), $H_{\text{eff}}(A,E)$. Given the generic form of the complexifier coherent states, however, this is a very hard task. For this reason (and to allow comparison with LQC effective dynamics), we limit ourselves to the homogeneous isotropic case. This is achieved by choosing very specific labels $h_e$ for our coherent states (\ref{coherent-state}): using the notation in (\ref{holomorphic-dec}), we choose $z_e = z := \xi + i\eta$ a complex number for all $e$ and $n_e=n_e'=n_{(i)}$ with $n_{(i)}$ being the $SU(2)$ element that rotates the $z$-axis into the tangent $\dot e$ to edge $e$. As one can check \cite{DL17b}, these states are indeed peaked on the elementary operators:
\begin{align}
\langle \hat{h}_{mn}(e)\rangle & = D^{(\frac{1}{2})}_{mn}(n_ee^{-i\xi\sigma_3/2}n_e^{\dagger}) [1 + \mathcal{O}(t)] \label{hol-peak}
\\
\langle \hat{E}^k(e)\rangle & = a^2\beta\eta D_{-k0}^{(1)}(n_e) [1+\mathcal O(t)] \label{flux-peak}
\end{align}
and their spread goes like $t$ (no summation over repeated indices)
\begin{align}
\langle \hat{h}_{mn}\hat{h}_{mn}\rangle - \langle\hat{h}_{mn}\rangle^2 & = \mathcal{O}(t) \label{hol-spread}
\\
\langle \hat{E}^k\hat{E}^k\rangle-\langle\hat{E}^k\rangle^2 & = \mathcal{O}(t) \label{flux-spread}
\end{align}
This sharp peakedness also allows us to deal with the {\it Gauss-constraint} of general relativity: since we are interested in observables which are themselves gauge-invariant (i.e., $U(g^{\dagger})\hat{H}U(g)=\hat{H}$ for any gauge transformation $U(g)$), and since the coherent states are gauge-covariant (i.e., $U(g) \psi_{e,h_e} = \psi_{e, h_e g^\dag}$), then one can see that group averaging produces
\begin{align}
\int dg \int dg' \langle \psi_{h_e}| U(g)^{\dagger}\hat{H}U(g')|\psi_{h_e}\rangle \approx \langle \psi_{h_e}|\hat{H}|\psi_{h_e}\rangle
\end{align}
with an error of order $\mathcal O(t)$. (Further details can be found in \cite{BT07,BT07_2}). This guarantees that our results have physical significance without having to solve the Gauss-constraint.

Looking at (\ref{hol-peak}) and (\ref{flux-peak}), we see that they coincide with the holonomy and flux computed in a classical flat Robertson-Walker spacetime described by $(p,c)$ if we perform the identification ($\mu\in\mathbb{R}$)
\begin{align}
\xi=\mu c,\hspace{20pt} \eta=\frac{\mu^2 p}{a^2\beta}
\end{align}
Under this identification, one also finds the following result for the expectation value of volume:
\begin{align} \label{volume-result}
\langle \hat V \rangle = \mathcal{N}^3 \mu^3 p^{3/2}[1+\mathcal{O}(t)]
\end{align}
which confirms that we should set $\mathcal N \mu = 1$ (so that the physical volume of the universe is given by $p^{3/2}$, the classical result). Under this identification, the expectation value of Hamiltonian is found to be (details of this computation will be in the companion paper to appear soon)
\begin{align} \label{hamiltonian-result}
& H_{\text{eff}}(p,c) := \langle \hat H \rangle = \langle \hat{H}_E \rangle - (\beta^2+1) \langle \hat{H}_L \rangle = \notag
\\
& = -\dfrac{6}{\kappa\beta^2} \sqrt p \dfrac{\sin(c\mu)^2}{\mu^2} [1 - (1 + \beta^2) \sin(c\mu)^2 + \mathcal{O}(t)]
\end{align}
where we used
\begin{align} \label{hamiltonian-result2}
\begin{array}{rl}
\langle \hat{H}_E\rangle & = \dfrac{6}{\kappa} \sqrt p \dfrac{\sin(c\mu)^2}{\mu^2} + \mathcal{O}(t)
\\
\langle \hat{H}_L \rangle & = \dfrac{6}{\kappa \beta^2} \sqrt p \dfrac{\sin(2c\mu)^2}{4\mu^2} + \mathcal{O}(t)
\end{array}
\end{align}
We notice several things: (1) at leading order in $\mu$ and $t$, $H_{\text{eff}}$ is consistent with the result obtained in LQC, equation (\ref{effective-rw-ham}); (2) in the classical limit ($t\to 0$) and continuum limit ($\mu \rightarrow 0$), $H_L$ and $H_E$ coincide (up to a numerical factor), and one recovers classical cosmology, equation (\ref{classical-rw-ham}); (3) if $\mu \neq 0$, however, the functional form of $H_L$ is different from that of $H_E$ in regard of the $c$-dependence, due to the different factor in the argument of sine. This is a modification to the established effective LQC Hamiltonian. A corresponding modification to the quantum LQC Hamiltonian $\hat H_{LQC}$ leads to a difference equation of higher order, whose quantum dynamics has been numerically investigated in \cite{ADLP}. It should be noted that the Hamiltonian in (\ref{hamiltonian-result}) corresponds to a modification of the effective LQC Hamiltonian in the so-called $\mu_0$-scheme.\footnote
{It might be possible to obtain the $\bar{\mu}$-scheme following the same procedure developed in \cite{AC16_mubar}.}
\section{A Toy Model}
What we derived in (\ref{hamiltonian-result}) and (\ref{hamiltonian-result2}) is the expectation value of the gravitational part of the LQG scalar constraint on a family of complexifier coherent states adapted to homogeneous isotropic cosmology. At this point it is not yet clear whether it serves as an effective Hamiltonian for the dynamics of full LQG, however it represents a promising candidate. We will now use it as an effective Hamiltonian on the phase space parametrized by $(p,c)$ and study the dynamics it produces. To facilitate the comparison with LQC effective dynamics, we choose as matter content a massless scalar field $\phi$, so that the effective scalar constraint reads
\begin{align}
H = H_{\text{eff}} + H_\phi, \ \ \ \ \ H_\phi = \dfrac{\pi_\phi^2}{2p^{3/2}}
\end{align}
with $\pi_\phi$ the momentum of $\phi$. The evolution of any phase space function $f$ wrt cosmological time $t$ is now obtained by Hamilton's equation $\dot f = \{f, H\}$, and in particular we see that $\pi_\phi$ is a constant of motion. We can now compute the system of $\dot p$ and $\dot c$, which can be numerically solved, leading to the phase space trajectories labelled by the constant of motion $\pi_\phi$. The evolution of volume $v = p^{3/2}$ wrt cosmological time $t$ and physical time $\phi$ is plotted in figure 1 for a simple example.
\begin{figure}
\begin{centering}
\includegraphics[height=7.5cm]{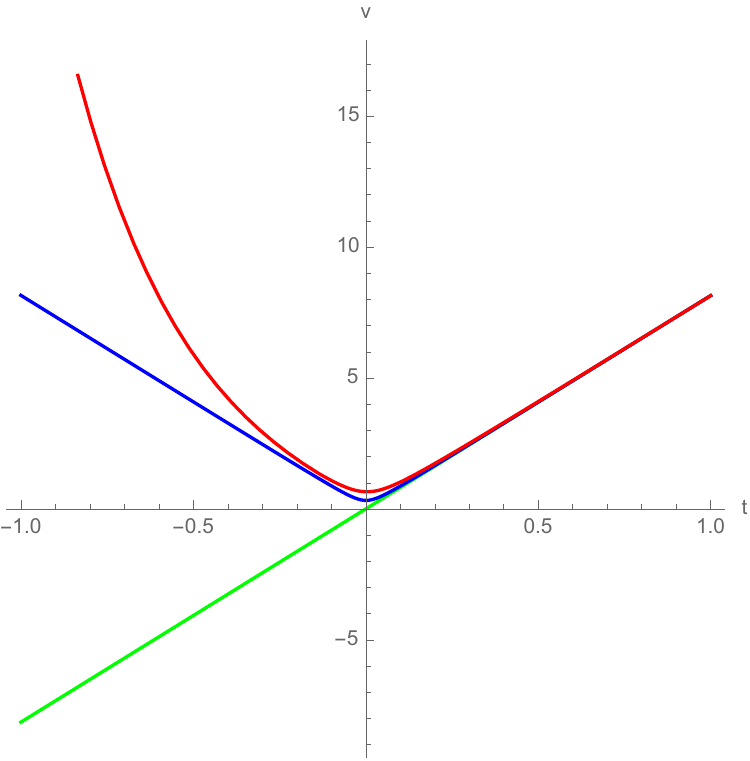}
\par\end{centering}
\begin{centering}
\includegraphics[height=7.5cm]{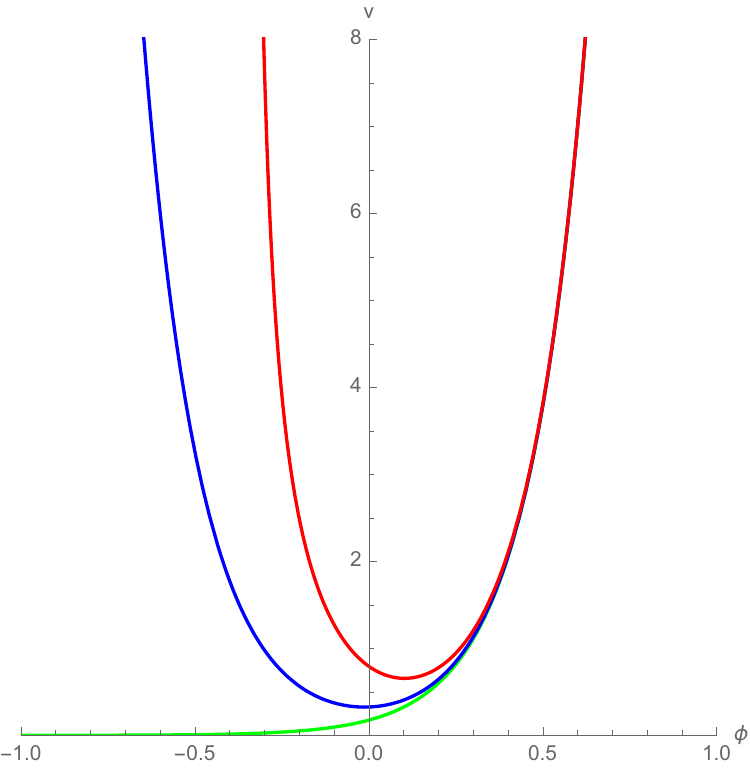}
\par\end{centering}
\caption{\footnotesize Plots of $v(t)$ and $v(\phi)$ in our toy model (red), LQC (blue) and classical cosmology (green, extended beyond the singularity to negative values for clarity), for $\pi_\phi = 1$ and $\beta = 0.12$.}
\end{figure}

As the plots show, the behavior of the universe at late time is the same as in LQC (and classical cosmology), but is quite different in the early universe: while a bounce still occurs, in our model it is highly non-symmetric and in the far past the universe does not obey classical Friedmann equations for a contracting spacetime. This is a major departure from the standard big bounce picture of LQC, and it is entirely due to the introduction of the Lorentzian term coming from the full theory.\footnote
{
A similar Hamiltonian has also been found within the symmetry-reduced case in \cite{yongge}, though the authors of that work did not emphasize the departure from the bouncing scenario of LQC that the new term entails. On the other hand, a similar modification of the bounce is found in the context of minisuperspace models with self-dual variables \cite{edward}. The similarity, however, is only superficial, since the effective Hamiltonian there is different from (\ref{hamiltonian-result}), and it is obtained by considering immaginary Immirzi parameter (for which LQG is not well-defined).
}

We should also mention that, while the Hamiltonian formulation of this problem is clear, it is not straightforward to recast it in terms of Friedmann and Raychaudhuri equations. This is due to the fact that the constraint equation $H = 0$ admits two solutions for $\sin(c\mu)^2$ as a function of $p$. For a detailed discussion, see \cite{param-new}.
\section{Conclusions}
The idea to obtain an effective scalar constraint (and hence effective dynamics) from the quantum theory has a long road of success. In the seminal paper \cite{APS}, the effective Hamiltonian was shown to reproduce the quantum evolution of coherent states, in particular removing the cosmologic singularity. First steps of repeating the effective dynamics program in a more complicated setting were taken in \cite{AC14_summary}, in the context of a gauge-fixed version of the full theory (called Quantum Reduced Loop Gravity, QRLG). In those works, the authors compute the expectation value of the QRLG Hamiltonian on some semiclassical states, and find that it coincides with the effective Hamiltonian (\ref{effective-rw-ham}) of LQC.

In this paper, we have pushed the program further, by calculating the expectation value of the LQG Hamiltonian, without assuming any simplifications, and considering both Euclidean and Lorentzian contributions. We found that the Euclidean part is in agreement with LQC at leading order (the corrections are proportional to the semiclassicality parameter $t$ of the complexifier coherent states we used). The Lorentzian part, on the other hand, introduces a different dependence on the variable $c$. While this dependence was already discussed within LQC \cite{yongge}, it was not emphasized that it leads to a departure from the standard bounce picture. We have shown that this is the case in a simple toy model.

The main message we want to communicate is twofold. On one hand, we have seen how from the full theory one can derive a candidate for an effective Hamiltonian of a reduced symmetry model (encoded in the special choice of coherent state labels), and that this is in agreement with LQC (and classical cosmology) at late times, whereas important modifications appear in the early universe. For applications in LQC the name ``effective Hamiltonian'' is indeed justified, as its trajectories agree with those obtained via quantum evolution of the minisuperspace model \cite{ADLP}.

On the other hand, we have shown that the choice of coherent states and the regularization procedure by which one obtains the Hamiltonian operator have crucial impact even on the semiclassical dynamics. Unless these details are fixed, no reliable quantitative and finer qualitative predictions can be made. The regularisation considered here and the other proposals will in general give different results. They are only comparable a posteriori, i.e., after physical quantities have been computed by methods such as the one presented in this paper. In lack of experimental evidence, this presents the necessity to find other methods by which the family of possible regularisations can be restricted. It is thus inescapable to integrate methods (such as the renormalization group \cite{Rg1} or consistency with Dirac algebra \cite{Bojo}) into the quantization process in order to fix any discretization errors.

\textbf{Acknowledgements.} The authors wish to thank Wojciech Kami\'nski, Marcin Kisielowski, Alexander Stottmeister, Thomas Thiemann, Kristina Giesel and Almut Oelmann for fruitful discussions. KL thanks the German National Merit Foundation for their financial support.


\begin{thebibliography}{99}

\bibitem{BCT11}
	M. Bojowald, G. Calcagni and S. Tsujikawa.
	{\it Phys. Rev. Lett.}
	{\bf 107}
	211302
	(2011)

%*************LQG

\bibitem{Rov04}
	Carlo Rovelli.
	Quantum Gravity.
	{\it Cambridge University Press}
	(2004)
	
\bibitem{Thi07}
	Thomas Thiemann.
	Modern Canonical Quantum General Relativity.
	{\it Cambridge University Press}
	(2007)
	
\bibitem{GS13}
	K. Giesel and H. Sahlmann.
	(2013)
	[arXiv:1203.2733v2]

%*************LGC
		
\bibitem{Boj08}
	Martin Bojowald.
	Loop Quantum Cosmology.
	(2008)
	{\it Living Reviews in Relativity}
	11: 4. doi:10.12942/lrr-2008-4
	
\bibitem{AS11}
	A. Ashtekar and P. Singh.
	\textit{Class. Quant. Grav.}
	\textbf{28}
	(2011)
	
\bibitem{AS16}
	I. Agullo and P. Singh.
	(2016)
	[arXiv:1612.01236]
	
\bibitem{GT15}
	K. Giesel and T.Thiemann.
	{\it Class. Quant. Grav.}
	\textbf{32}
	(2015)

\bibitem{GO17}
	K. Giesel and A. Oelmann.
	(2016)
	[arXiv:1610.07422]

\bibitem{APS-PRL}
	A. Ashtekar, T. Pawlowski and P. Singh.
	{\it Phys. Rev. Lett.}
	{\bf 96}
	(2006)
	
\bibitem{APS}
	A. Ashtekar, T. Pawlowski and P. Singh.
	{\it Phys. Rev. D}
	\textbf{74}
	(2006)
	084003

\bibitem{Tav08}
	V. Taveras.
	{\it Phys. Rev. D}
	{\bf 78}
	(2008)



%**********Hamiltonian

\bibitem{Ash87}
	Abhay Ashtekar.
	In Mathematics and General Relativity 
	(American Mathematical Society, Providence, Rhode Island, 1987)
	
\bibitem{Ash88}
	A Ashtekar. 
	{\it Contemporary Math.}
	\textbf{71}
	(1988) 	

\bibitem{Bar94}
	J. Fernando Barbero.
	{\it Phys. Rev. D}
	\textbf{49}
	6935-6938
	(1994)
	
\bibitem{Bar95}
	J. Fernando Barbero.
	{\it Phys. Rev. D}
	\textbf{51}
	5507-5510
	(1995)
	
\bibitem{Bar96}
	J. Fernando Barbero.
	{\it Phys. Rev. D}
	\textbf{54}
	1492-1499
	(1996)
	
\bibitem{GT06-2}
	K. Giesel and T. Thiemann.
	\textit{Class.Quant.Grav.}
	\textbf{24}
	(2007)

\bibitem{Thi96_1}
	Thomas Thiemann.
	\textit{Class.Quant.Grav.}
	\textbf{15}
	(1998)
		
\bibitem{Thi96_2}
	Thomas Thiemann.
	\textit{Class.Quant.Grav.}
	\textbf{15}
	(1998)

\bibitem{AL97}
	A. Ashtekar and J. Lewandowski.
	\textit{Adv.Theor.Math.Phys.}
	\textbf{1}
	388-429
	(1998)
	
\bibitem{GT_CC06}
	K. Giesel and T. Thiemann.
	{\it Class. Quant. Grav.}
	\textbf{23}
	(2006)

%******************Coherent States
	
\bibitem{Thi00_I}
	T. Thiemann,
	\textit{Class. Quant. Grav.}
	\textbf{18}
	2025-2064
	(2000)
	
\bibitem{TW01_II}
	T. Thiemann, O. Winkler.
	\textit{Class. Quant. Grav.}
	\textbf{18}
	(2001)

\bibitem{TW01_III}
	T. Thiemann and O. Winkler.
	\textit{Class.Quant.Grav.}
	\textbf{ 18}
	4629-4682
	(2001)

\bibitem{Thi02}
	Thomas Thiemann.
	\textit{Class.Quant.Grav.}
	\textbf{ 23}
	(2006)
	2063-2118
	
\bibitem{Hall94}
	B.C. Hall.
	{\it Journ. Funct. Analysis}
	\textbf{122}
	103-151
	(1994)
	
\bibitem{Hall97}
	B.C. Hall.
	{\it Comm. Math. Phys.}
	\textbf{184}
	233-250
	(1997)
	
%**********Volume Operator**************

\bibitem{GT06}
	K. Giesel and T. Thiemann.
	\textit{Class.Quant.Grav.}
	\textbf{24}
	(2006)
	
	
\bibitem{DL17b}
	A. Dapor and K. Liegener.
	(2017)
	[arXiv:1710.04015]
	
%*********************Gauss-Constraint********

\bibitem{BT07}
	B. Bahr and T. Thiemann.
	\textit{Class.Quant.Grav.}
	\textbf{26}
	(2009)
	
\bibitem{BT07_2}
	B. Bahr and T. Thiemann.
	\textit{Class.Quant.Grav.}
	\textbf{26}
	(2009)

%****************Quantum Reduced Loop Gravity and Yongge

\bibitem{ADLP}
	M. Assanioussi, A. Dapor, K. Liegener, T. Pawlowski
	(2018)
	[arXiv:1801.00768]
	
\bibitem{AC16_mubar}
	E. Alesci and F. Cianfrani.
	(2016)
	[arXiv:1604.02375]	
	
\bibitem{yongge}
	Y. Ding, Y. Ma. and J. Yang
	\textit{Phys.Rev.Lett.}
	\textbf{102}
	051301
	(2009)

\bibitem{edward}
	E. Wilson-Ewing
	\textit{Phys. Rev. D}
	\textbf{92}
	123536
	(2015)

\bibitem{param-new}
	B.-F. Li, P. Singh, A. Wang
	(2018)
	[arXiv:1801.07313]

\bibitem{AC14_summary}
	E. Alesci and F. Cianfrani.
	(2014)
	[arXiv:1410.4788]\\
	E. Alesci and F. Cianfrani.
	(2014)	
	[arXiv:1402.3155]\\
	E. Alesci and F. Cianfrani.
	(2015)
	[arXiv:1506.07835]
	
% ***** Renormalisation ***

\bibitem{Rg1}
	T. Lang, K. Liegener, T. Thiemann.
	(2018)
	[1711.05685]\\
	T. Lang, K. Liegener, T. Thiemann.
	(2018)
	[1711.06727] \\
	T. Lang, K. Liegener, T. Thiemann.
	(2018)
	[1711.05688]\\
	T. Lang, K. Liegener, T. Thiemann.
	(2018)
	[1711.05695] 

% ***** Consistent-Algebra ***

\bibitem{Bojo}
	M. Bojowald, S. Brahma, D. Yeom.
	(2018)
	[1803.01119]	
	
	
	
\end{thebibliography}
\end{document}